\documentclass[a4paper,12pt]{article}

\usepackage[english]{babel}
\usepackage[utf8x]{inputenc}
\usepackage[T1]{fontenc}
\usepackage{authblk}

\usepackage[margin=2.5cm,nohead]{geometry}

\usepackage{amsmath}
\usepackage{amsthm}
\usepackage{graphicx}
\usepackage{multirow}
\usepackage[colorinlistoftodos]{todonotes}
\usepackage[colorlinks=true, allcolors=blue]{hyperref}
\usepackage[comma]{natbib}
\usepackage{float}
\usepackage{setspace}
\usepackage{color} 
\title{Expropriations, Property Confiscations and New Offshore Entities: Evidence from the Panama Papers\thanks{We thank Ulrich Matter for helpful comments. Ralph-C.\ Bayer, Roland Hodler and Paul Raschky gratefully acknowledge financial support from the Australian Research Council (ARC Discovery Grants DP12010183 and DP150100061).}}

\author{Ralph-C.\ Bayer\thanks{School of Economics, University of Adelaide. Email: ralph.bayer@adelaide.edu.au} \hspace{1cm} Roland Hodler\thanks{Department of Economics, University of St. Gallen; CEPR; CESifo. Email: roland.hodler@unisg.ch.} \\ Paul A.\ Raschky\thanks{Department of Economics, Monash University. Email: paul.raschky@monash.edu.} \hspace{1cm} Anthony Strittmatter\thanks{Department of Economics, University of St. Gallen. Email: anthony.strittmatter@unisg.ch.}}

\begin{document}
\maketitle

\begin{abstract}
\noindent Using the Panama Papers, we show that the beginning of media reporting on expropriations and property confiscations in a country increases the probability that offshore entities are incorporated by agents from the same country in the same month. This result is robust to the use of country-year fixed effects and the exclusion of tax havens. Further analysis shows that the effect is driven by countries with non-corrupt and effective governments, which supports the notion that offshore entities are incorporated when reasonably well-intended and well-functioning governments become more serious about fighting organized crime by confiscating proceeds of crime. \bigskip

 \bigskip

\end{abstract}

\newpage

%%%%%%%%%%%%%%%%%%%%%%%%%%%%%%%%%%%%%%%%%%%%%%%%%%%%%%%%%%%%%%%%%%%%%%%%%%%%%%%%%%%%
\section{Introduction}

The wealth hidden in offshore entities in tax havens has become a hotly debated topic. The recent academic work by \cite{zucman2013missing,zucman2015hidden} and the leaking of the so-called Panama Papers have been instrumental in the increased attention the topic has received. The Panama Papers contain information on more than 200,000 offshore entities and are in a de-identified form in the public domain due to a leak at the Panamanian law firm and corporate service provider Mossack Fonseca \& Co. From a policy perspective, it is important to understand why individuals and firms decide to hide their wealth in offshore entities. Most certainly, the firms and individuals compare the expected costs and benefits from hiding and not hiding their wealth. Many studies (reviewed below) thus focus on tax minimization as the driver behind offshore entities. 

A so far little discussed motivation is the fear of expropriation. In many countries, individuals and firms face the risk of various forms of expropriation. Depending on the governance in a country, different groups might be affected. In countries with good governance, the fear is mainly with the bad guys while in countries with bad governance, the good guys are under threat. For countries with benevolent governments and well-functioning law enforcement, the confiscation of proceeds from and assets used in crime is an important deterrence and enforcement tool. It is a plausible hypothesis that criminal individuals and organizations move assets offshore when their perceived risk of expropriation increases. Similarly, productive respectable citizens and companies residing in weakly institutionalized countries with Leviathan governments or thieving politicians could reasonably be expected to react by shifting assets offshore when the fear of expropriation increases. Our study aims at understanding whether changes in the perceived risk of expropriation and property confiscation can induce individuals and firms to incorporate offshore entities.

We use information on the incorporation of offshore entities from the Panama Papers, as well as information on news reports on expropriations and property confiscations from the GDELT Project. The underlying idea is that such news reports induce private individuals and organizations to update their beliefs about the expropriation risk by the government. Hence, for some individuals it may become optimal to incorporate an offshore entity and to transfer their wealth offshore. Using a sample of 160 countries and monthly observations from 2007 to 2012, we find that the beginning of a spell of news reports on expropriations and property confiscations in a country increases the probability that an offshore entity is incorporated by an agent from the same country in the same month by around three percentage points. This result is robust to the use of country-year fixed effects, which control for all characteristics that are constant within any country and year, and the exclusion of tax havens as countries of origin.

In the next step, we investigate whether this average effect is driven by shady figures and organizations fleeing potential confiscation by well-intended governments, or by respectable individuals protecting their wealth from a Leviathan government that has revealed its malevolence through previous property expropriation that was featured in the news. %To put it more succinctly, we ask whether it is bad guys hiding wealth from good governments, or good guys hiding wealth from bad governments. 
To address this question, we split the sample using various measures of governance and institutional quality (and economic development), and reestimate our main specification for the respective subsamples. We find that the positive effect of the beginning of a spell of news reports on expropriations and property confiscations on the incorporation of offshore entities remains large and statistically significant for (rich) countries with non-corrupt and effective governments. These results support the notion that criminals and criminal organizations use offshore entities to protect their illegally acquired wealth from reasonably well-intended and well-functioning government agencies. In contrast, we find no statistically significant effects in countries with poor governance. Hence, we provide no evidence for the notion that honest individuals and firms use offshore entities to protect their wealth from Leviathan governments.\footnote{We discuss possible reasons for this negative result in Section 5.} 

Our study is related to the extensive literature on tax evasion and, more specifically, to contributions on tax evasion through the use of tax havens \citep[e.g.,][]{allingham1972income,klepper1989tax,slemrod2007cheating,gravelle2009tax,hin10,davies2018,johannesen2018}. These contributions focus, among other things, on how tax evasion and offshore sheltering depends on the fiscal environment, including tax rates, enforcement and legal consequences of detection. \cite{bennedsen2018} show that sometimes managers use complex nets of offshore entities to cover embezzlement of company funds. \cite{Desai2006} and \cite{Gumpert2016} identify characteristics of American and German firms that use offshore affiliates for tax purposes. We focus on a different driver of demand and investigate whether the perceived risk of expropriation and property confiscation (and changes thereof) has an impact on offshore activity. 

We are not the first to make use of the Panama Papers. \cite{alstadsaeter2017tax} estimate the size and distribution of tax evasion in Sweden and document that the wealthiest individuals evade a much higher share of their personal taxes than the average citizen. \cite{alstadsaeter2017owns} take this approach to the global level and show that wealth inequality is much higher than measures computed with tax data suggest.
Building on earlier work by \cite{johannesen2014tax}, the studies by \cite{caruana2016offshore} and \cite{omartiantax} use the Panama Papers to study the effectiveness of the European Savings Directive and other policies aimed at fighting tax evasion. In a line of research complementary to the focus on tax evasion and our focus on the risk of expropriations and property confiscations, \cite{andersen2017petro} and the International Consortium of Investigative Journalists (ICIJ), which made the Panama Papers publicly available, focus on the use of offshore entities by political leaders and public officials.\footnote{The ICIJ finds 246 offshore entities with a direct or indirect connection to political leaders or public officials, which corresponds to 0.1\% of all the offshore entities in the Panama Papers. For more information on their investigation, see \url{https://panamapapers.icij.org/the_power_players/}.} 

This paper also relates to the literature on corporate strategy in the presence of appropriation risk. \cite{caprio2013shelter} use data on more than 30,000 publicly traded firms from 109 countries and find that firms located in countries with a higher threat of political extraction have lower holdings in liquid assets and have higher investments in fixed assets. The intuition is that cash and other liquid holdings are more susceptible to the greedy hand of corrupt politicians compared to fixed assets. Similarly, firms conduct less direct investment in countries with high expropriation risk \citep[e.g.,][]{Azzimonti2018}.

Findings from cross-country studies further suggest that in countries with more extractive or corrupt political institutions firms avoid the risk of expropriation by going underground \citep{johnson1988under} or having a more concentrated corporate ownership structure \citep{stulz2005owner}. Our study complements this literature by focusing on off-shoring assets as another channel of avoiding the risk of expropriation. 

While lawful organizations might react to the risk of being expropriated by corrupt governments, organizations involved in illegal activity (such as drug or arms trafficking) in many countries face the risk of confiscation of proceeds from crime and the freezing of assets. \cite{bowles2000,bowles2005} conduct an economic analysis of forfeiture laws and conclude that they are a useful complement to other enforcement instruments. Reliable evidence for the effect of confiscation laws is not readily available. To our knowledge, we are the first to demonstrate that the perceived risk of confiscation of proceeds from crime leads to an increase in offshore activity.

The remainder of this paper is structured as follows: Section 2 provides a simple conceptual framework for thinking about a private individual's decision about whether to hide her wealth in an offshore entity. Section 3 introduces the data, and Section 4 the empirical specification. Section 5 presents our results. Section 6 briefly concludes.

%%%%%%%%%%%%%%%%%%%%%%%%%%%%%%%%%%%%%%%%%%%%%%%%%%%%%%%%%%%%%%%%%%%%%%%%%%%%%%%%%%%%
\section{Decision to hide wealth offshore}

In this section, we present a simple decision problem of an individual who (i) needs to decide whether to hide her wealth offshore and (ii) may be expropriated by the government if she decided against hiding her wealth offshore. This simple decision problem may represent the situation of a shady individual facing a reasonably well-meaning and well-functioning government or the situation of an honest individual facing a Leviathan government.

For this individual $i$, hiding her wealth offshore is associated with transaction costs and perhaps some opportunity costs as her wealth cannot be invested elsewhere, as well as the risk that she may not be able to repatriate her wealth later on. We simply denote her expected (net) benefit from hiding her wealth offshore by $W^o_i$. Similarly, we denote by $W^n_i$ her expected benefit from not hiding her wealth if the government does not expropriate her. Finally, we denote by $\mu_i$ her belief that the government expropriates her and confiscates her assets if she does not transfer them offshore. This individual decides to hide her wealth offshore if and only if
\begin{equation}
W^o_i>(1-\mu_i)W^n_i \,\, \Leftrightarrow \,\, \mu_i>\frac{W^n_i-W^o_i}{W^n_i}\equiv\hat\mu_i, 
\end{equation}
where $\hat\mu_i$ is an (individual-specific) threshold. 

Individual $i$'s belief $\mu_i$ may differ from the true probability that the government expropriates her, and she may be aware of the true probability. Thus, she may use additional information, such as a news report on expropriations and property confiscations, to update her belief about the government's type (e.g., how seriously the government enforces the rules or how aggressively it expropriates wealthy individuals). More importantly, such a news report may increase her belief $\mu_i$ that the government will expropriate her if she does not hide her wealth offshore. If $\mu_i$ was below $\hat\mu_i$ prior to the news report but is above it thereafter, then the news report induces the individual to hide her wealth offshore.

\section{Data description}

We collect data from various sources to collect a balanced monthly panel for 160 countries and the years 2007--2012. The main variables are the incorporation of offshore entities based on the Panama Papers and news reports on expropriations and property confiscations from the GDELT data. 

%%%%%%%%%%%%%%%%%%%%%%%%%%%%%%%%%%%%%%%%%%%%%%%%%%%%%%%%%%%%%%%%%%%%%%%%%%%%%%%%%%%%
\subsection{Panama Papers and the incorporation of offshore entities}

The International Consortium of Investigative Journalists (ICIJ) obtained data on around 214,000 offshore entities due to a leak at the Panamanian law firm and corporate service provider Mossack Fonseca \& Co.\footnote{The ICIJ provides the data at \url{https://offshoreleaks.icij.org/pages/database}. We retrieved the data in 
May 2016.} These data became known as the so-called Panama Papers. Offshore entities are companies, trusts, or funds registered in so-called tax havens, i.e., low-tax jurisdictions. Most of the offshore entities in the Panama Papers have jurisdiction in the British Virgin Islands (53\%) or in Panama (23\%). Many others have jurisdiction in the Bahamas (7\%), the Seychelles (7\%) and Niue (5\%).\footnote{Table \ref{tab_entjur} in the Appendix lists the jurisdictions of all offshore entities in the Panama Papers.} The agents registering offshore entities can be natural or legal persons, and they are typically located in a country or jurisdiction outside the offshore jurisdiction. The Panama Papers include offshore entities registered by agents from 160 different countries and jurisdictions.

\begin{figure}[h!]
\begin{center}
\caption{Number of offshore entity incorporations by year.} \label{fig_incorpyear}
\includegraphics[width=11cm]{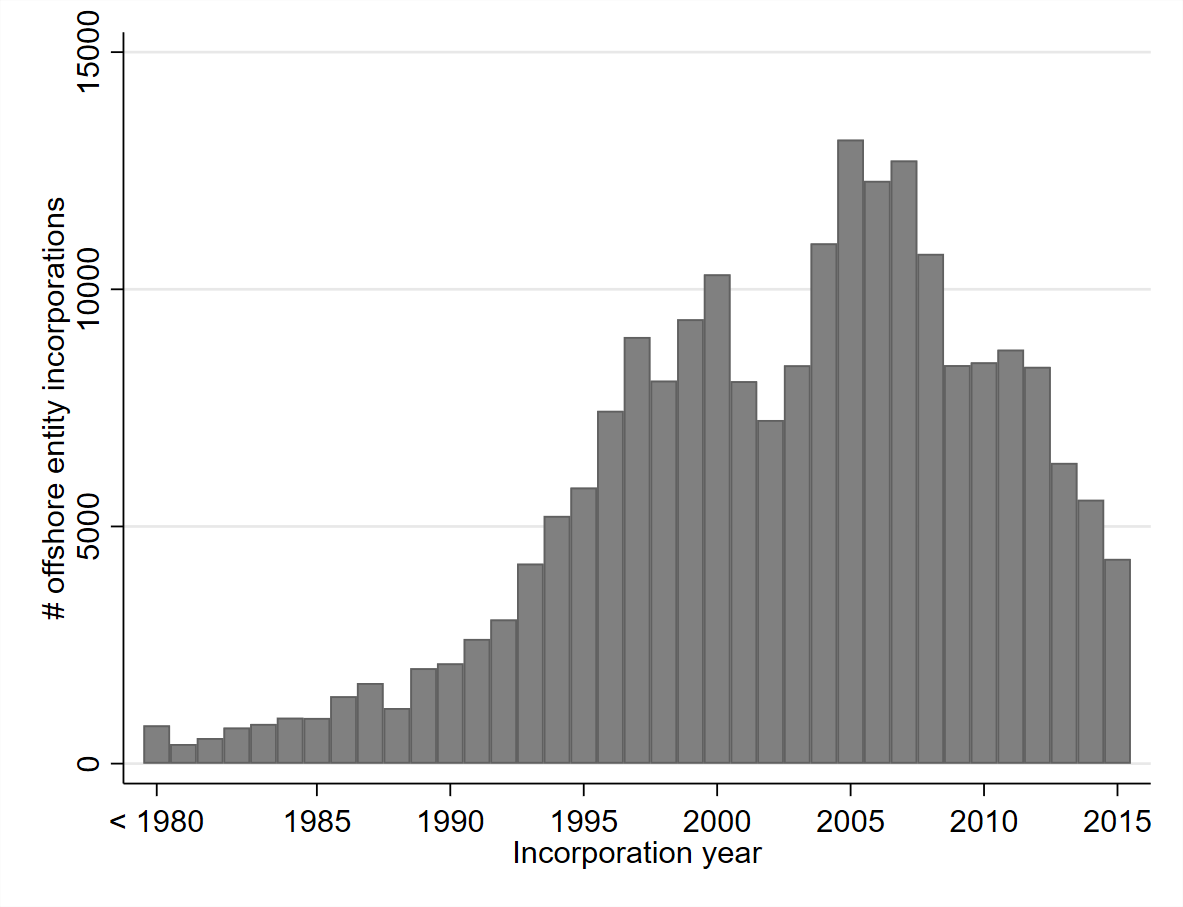}
\end{center}
\parbox{\textwidth}{\footnotesize Note: Own calculations based on data from the Panama Papers. }
\end{figure}

\begin{figure}[h!]
\begin{center}
\caption{Histogram of the number of offshore entity incorporations by country and month.} \label{fig_incorpyearmonth}
\includegraphics[width=11cm]{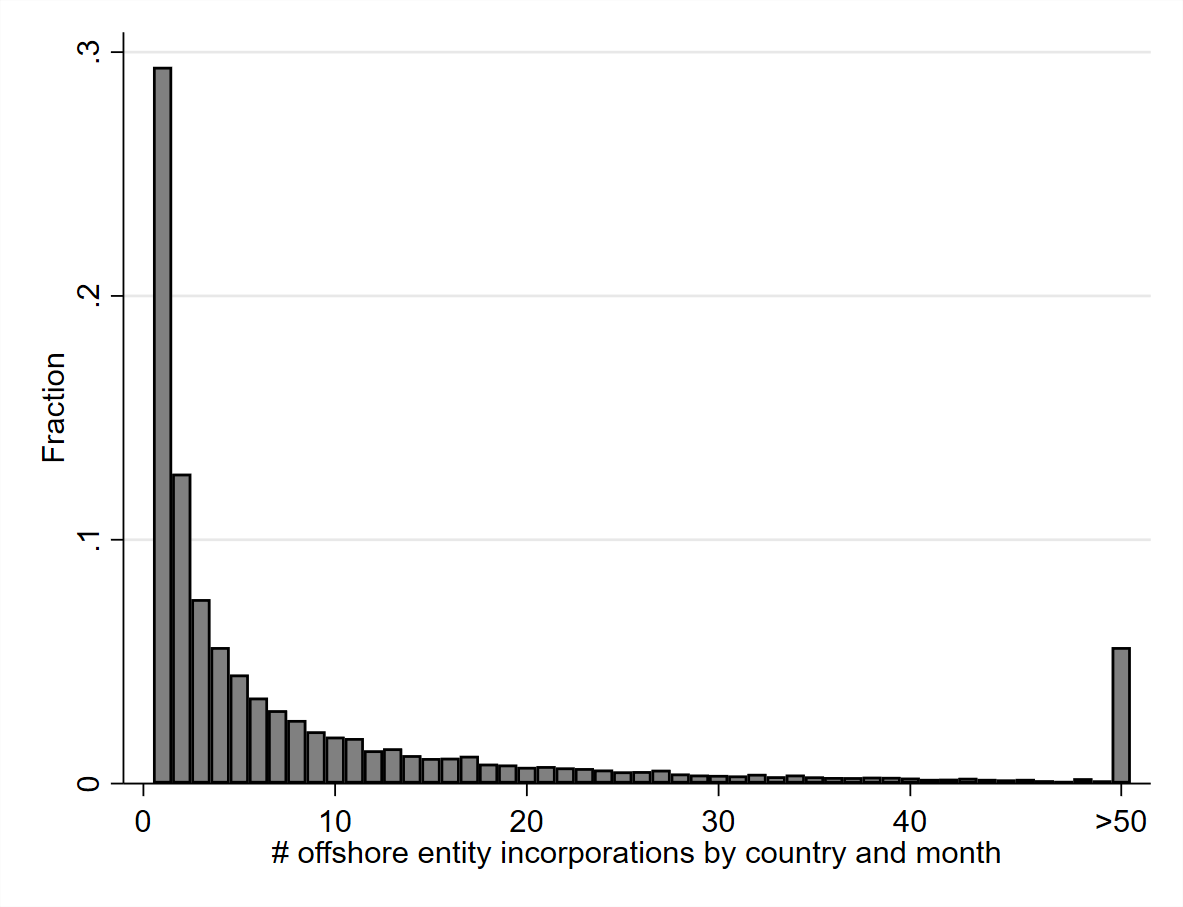}
\end{center}
\parbox{\textwidth}{\footnotesize Note: Own calculations based on data from the Panama Papers. We include only countries and months with a positive number of offshore entity incorporations.}
\end{figure}

The information provided by the Panama Papers includes detailed incorporation dates for the offshore entities.\footnote{We drop 816 offshore entities because of missing incorporation dates. We also drop the 246 offshore entities with a direct or indirect connection to political leaders or public officials according to the ICIJ. The list of these offshore entities is available upon request.} Figure \ref{fig_incorpyear} reports the observed number of incorporations of offshore entities by year. 
\[ \text{Figure \ref{fig_incorpyear} around here.} \]
There are more incorporations in the years 1996--2012 than in earlier or later years. The highest number of incorporations are from 2005 to 2007, with around 12,000 incorporations per year. 
Figure \ref{fig_incorpyearmonth} shows a histogram of the number of incorporations of offshore entities by the registering agents' country/jurisdiction and month. 
\[ \text{Figure \ref{fig_incorpyearmonth} around here.} \]
The distribution is highly right-skewed, with the mean number of incorporations per country and month being 13 and the median being 4. The maximum of 551 offshore entities was incorporated by Swiss agents in April 2005.

Our dependent variable, $\mbox{\textit{Offshore}}_{imy}$, is the binary variable indicating whether there is at least one incorporation of an offshore entity in a specific country/jurisdiction $i$ and month $my$ (with $m$ indicating the month of the year and $y$ the year). The use of a binary variable avoids dealing with the highly right-skewed number of incorporations, which include extreme outliers. In the raw data starting in 1980, the share of countries and months with at least one incorporation of an offshore entity is 19\%. For the time period 2007--2012 used in our analysis, the corresponding share is 27\%. 

Table \ref{tab_countries} in the Appendix reports the number of months with offshore entity incorporations by the country/jurisdiction of the registering agents. Offshore entities are most regularly registered by agents in Hong Kong, Jersey, Luxembourg and Panama. One possible concern is that agents from such tax havens may often register offshore entities for clients located elsewhere. If so, the connection between news reports on expropriations and property confiscation within a specific country and the incorporation of an offshore entity from an agent from the same country breaks down. For example, a Luxembourgian agent may register an entity in the British Virgin Islands for a German client. Although the German client's decision may have been driven by news reports in Germany, our reliance on the agent's country implies that we would link the incorporation of the corresponding offshore unit to news reports on Luxembourg. Therefore, we drop the countries classified as tax havens by \cite{hin10} and \cite{joh14} in most of our estimates. Following the definition of \cite{hin10}, tax havens provide low tax rates and favorable regulatory policies to foreign investors offshore. \cite{joh14} exploit that G20 countries compelled offshore countries to sign bilateral treaties providing for exchange of bank information. %\cite{kol15} define offshore countries by their population size (less than 250,000), thereby classifying fewer countries as tax havens that the two other studies. 
Table \ref{tab_taxhavens} in the Appendix lists the 53 countries or jurisdictions of our sample that have been classified as tax havens by at least one of these studies.

%%%%%%%%%%%%%%%%%%%%%%%%%%%%%%%%%%%%%%%%%%%%%%%%%%%%%%%%%%%%%%%%%%%%%%%%%%%%%%%%%%%%
\subsection{GDELT data and property confiscations}

%I decided to use only the years between 2000 and 2012. I dropped the 2013/03-2013/08 because of the reasons we discussed. But then I realized that there are also only few property confiscations in 2014. Then I decided to drop all observations in 2013 and 2014, such that we have a balanced number of months per year. However, the results do not change (much) when we include 2013 and 2014.\\
%The more serious problems are the years before 2007 (see Figure \ref{fig3}). I guess we have a data problem here (I don't see a reason why there should be so little property confiscations before 2007). However, this shouldn't matter much when we include time dummies. I estimate one specification with the years 2000-2012 and another one with the years 2007-2012. The results are robust. However, when we go further back to the years 1990-2012 the results become insignificant. This might be of reasons related with the data quality in the 90s?  

Our data on expropriations and property confiscations are sourced from the GDELT 1.0 Database. The GDELT Project collects daily news event information from ``the world’s broadcast, print, and web news from nearly every corner of every country in over 100 languages” \cite[][]{leetaru13}. News reports from other languages are translated into English through a collaboration with Google Ideas. Each news report is fed into a parsing algorithm that automatically extracts information about the time and location of the event, as well as classifies it into categories and defines the actors involved.%\footnote{We retrieved the GDELT data in September 2016. The data were available until the end of 2013, but the year 2013 contains many events without information on the event actors. Therefore, we omit the last available year from our analysis. }

The GDELT Project uses 20 main event classifications based on the Conflict and Mediation Event Observations (CAMEO) Event and Actor Codebook. For the construction of our main explanatory variable, $\mbox{\textit{Confiscation}}_{imy}$, we use all events from the CAMEO category ``1711: Confiscate property.’’ These events are characterized by the verbs ``[u]se force to take control of somebody else’s property, confiscate, expropriate.’’ For simplicity, we will often refer to these events as property confiscations, but one should keep in mind that they include expropriations. For each event, GDELT defines a source and a target actor. We exclude all events without information on the actors as well as events for which source and target actors are from different countries.\footnote{We retrieved the GDELT data in September 2016, when it included observations until the end of 2013. For most observations in 2013, GDELT did not contain any information on the event actors. Therefore, we omit the last available year from our analysis.}

A major advantage of the GDELT data is the high temporal resolution, which allows construction of monthly aggregates. The use of monthly rather than annual aggregates has at least two advantages. First, from a theoretical perspective, news reports on property confiscations may lead to immediate changes in beliefs and therefore, potentially to prompt decisions to incorporate an offshore entity. Such prompt effects are easier to capture using monthly rather than annual data. Second, from a methodological perspective, the use of monthly aggregates allows exploitation of within-year and country variation in property confiscations and the corresponding news reports. %Other daily event datasets (i.e. ACLED or UDCP) are often limited in geographic coverage do not contain information about property confiscation. 

\begin{figure}[h!]
\begin{center}
\caption{Number of reported property confiscations by year.} \label{fig_confyear}
\includegraphics[width=11cm]{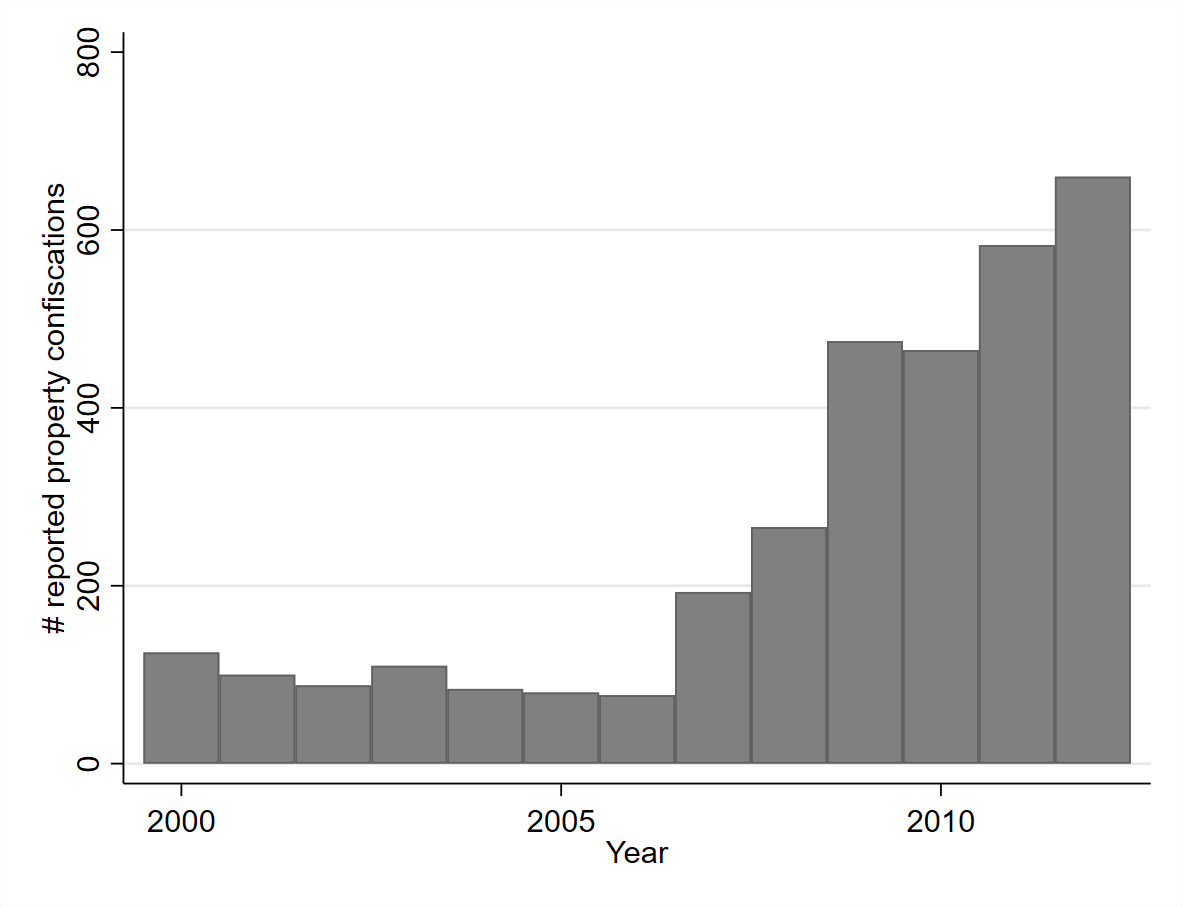}
\end{center}
\parbox{\textwidth}{\footnotesize Note: Own calculations based on data from GDELT from the years 2000-2012.}
\end{figure}

One caveat of the GDELT data is that they are sourced from online news reports and rely on automatic coding. As a result, data coverage and quality can vary over time. There are 3,306 reported property confiscations since 2000 in the GDELT data. Figure \ref{fig_confyear} shows the annual number of reported property confiscations for the years 2000--2012.
\[ \text{Figure \ref{fig_confyear} around here.} \]
There are only around 100 reported property confiscations per year before 2007. Then, there is a strong and steady increase from 2007 onward. The number of reported property confiscations peaks in 2012 with 661 confiscations. There could be different reasons for the low incidence of reported property confiscation before 2007. The main reason for this is that GDELT was less able to detect news reports on property confiscations in these early years than in later years, once improved algorithms were employed. Therefore, we focus on the years from 2007 to 2012 in our subsequent analysis.\footnote{We provide robustness tests using all years from 2000 to 2012 (see Table \ref{tab_mainlong} in the Appendix).} In addition, we will use various fixed effects to account for country- and time-specific shocks in data coverage and quality.

\begin{figure}[h!]
\begin{center}
\caption{Histogram of the number of reported property confiscations by country.} \label{fig_confcountry}
\includegraphics[width=11cm]{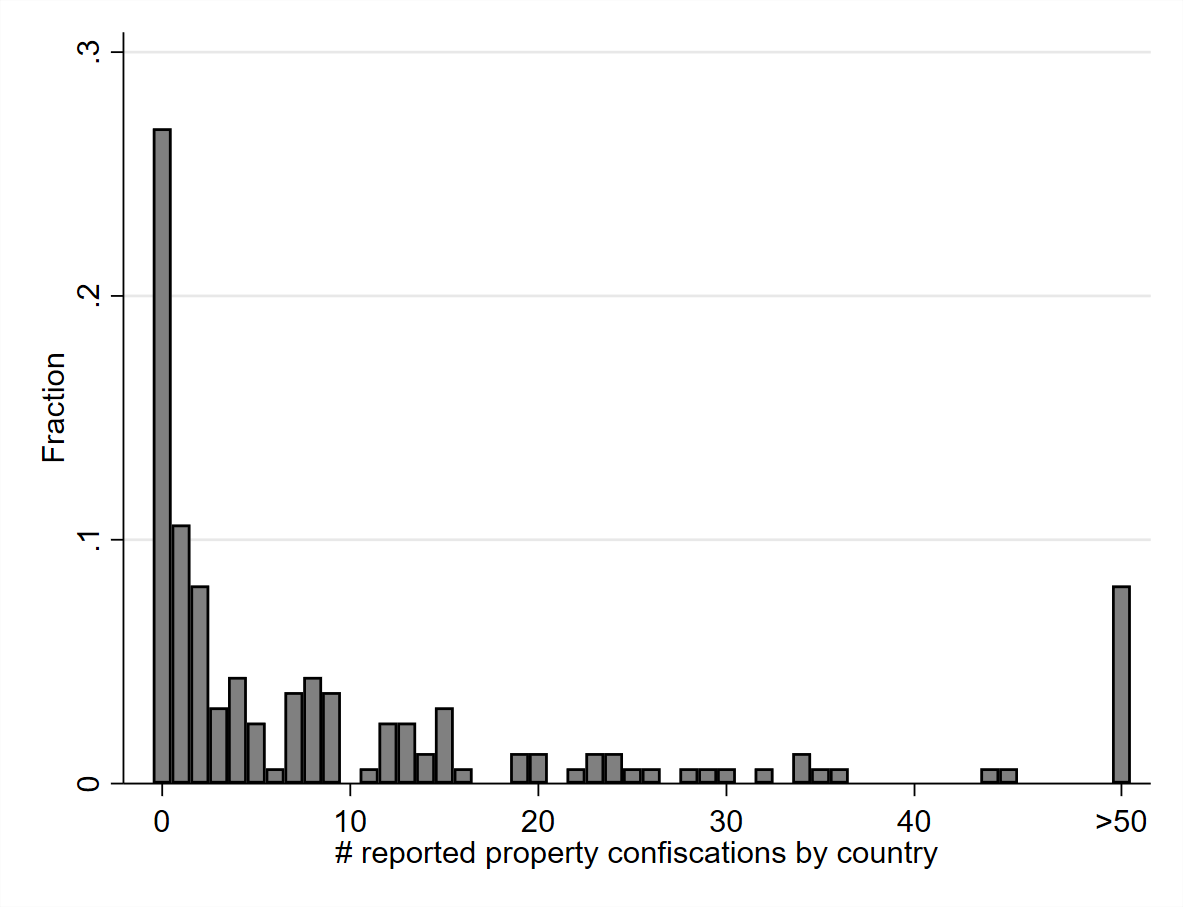}
\end{center}
\parbox{\textwidth}{\footnotesize Note: Own calculations based on data from GDELT from the years 2007-2012.}
\end{figure}

Figure \ref{fig_confcountry} documents the histogram of reported property confiscations by country.
\[ \text{Figure \ref{fig_confcountry} around here.} \]
The distribution is again right-skewed. There are no reports on property confiscations for 27\% of all the countries. The average number of reported property confiscations is 17, and the median is 4. The countries with the most reported property confiscations are Australia (135) and the United States (694).

\begin{figure}[h!]
\begin{center}
\caption{Histogram of the number of reported property confiscations by country and month.} \label{fig_confcountrymonth}
\includegraphics[width=11cm]{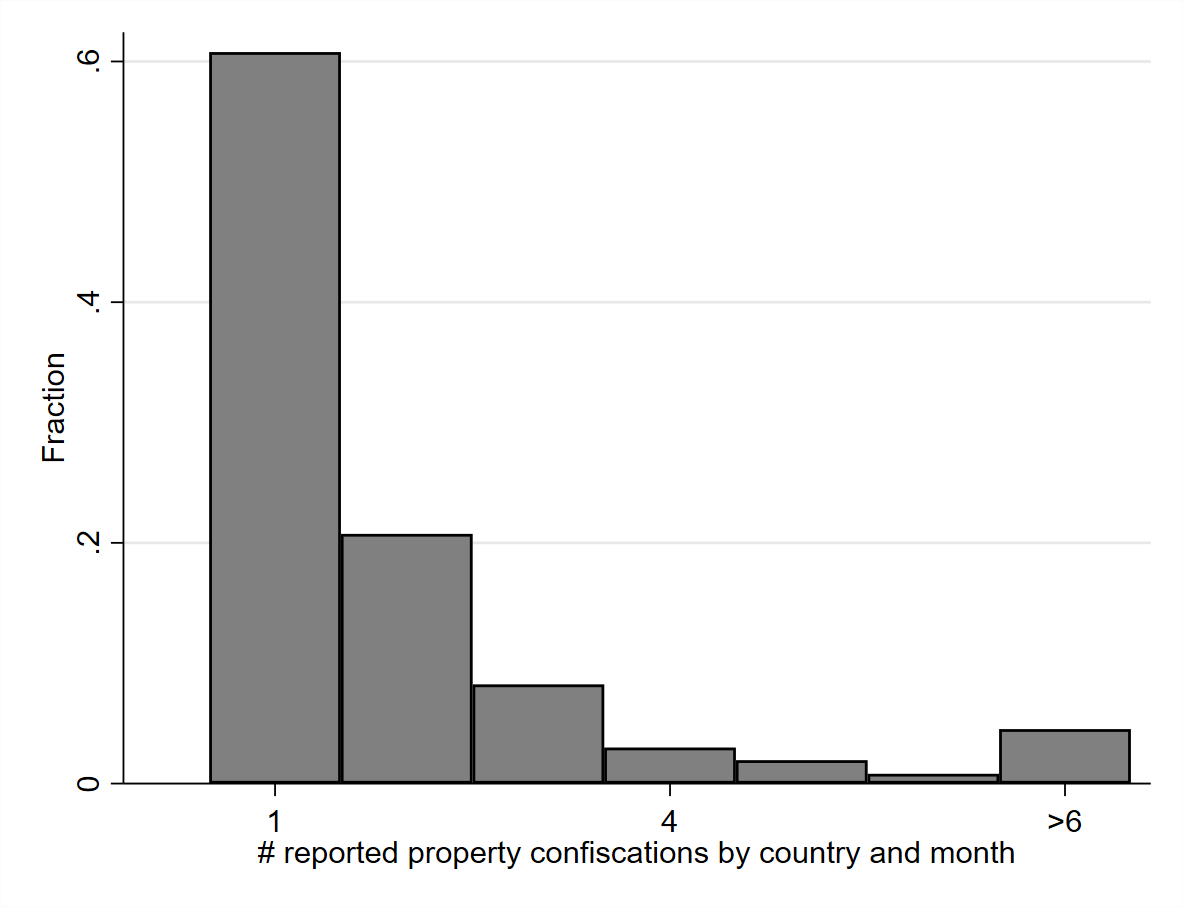}
\end{center}
\parbox{\textwidth}{\footnotesize Note: Own calculations based on data from GDELT from the years 2007-2012. We include only countries and months with a positive number of reported property confiscations.}
\end{figure}

\begin{figure}[h!]
\begin{center}
\caption{Histogram of the duration of property confiscating spells.} \label{fig_confspells}
\includegraphics[width=11cm]{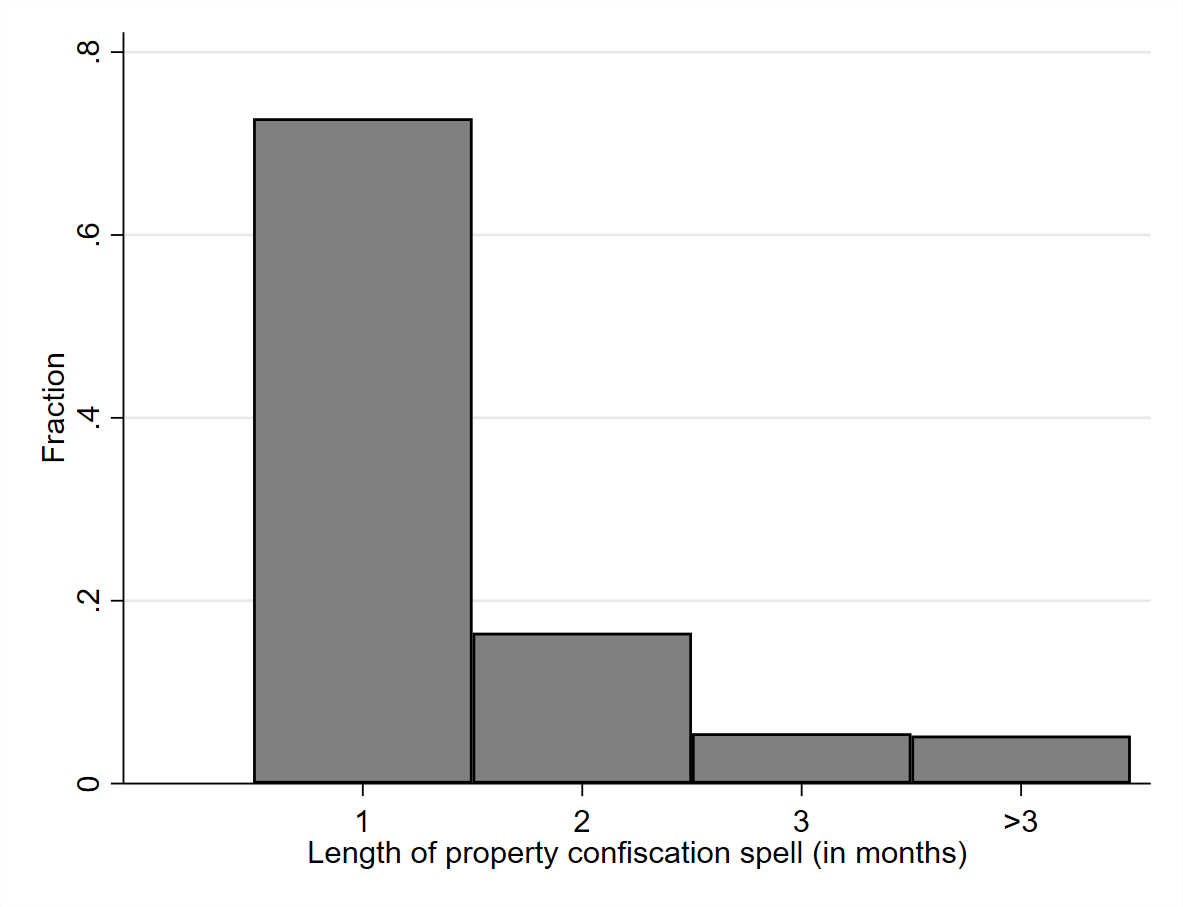}
\end{center}
\parbox{\textwidth}{\footnotesize Note: Own calculations based on data from GDELT from the years 2007-2012.}
\end{figure}

Many of the reported property confiscations happen in the same country and month, or in consecutive months within the same country. Figure \ref{fig_confcountrymonth} provides information on the number of reported property confiscations by country and month. 
\[ \text{Figure \ref{fig_confcountrymonth} around here.} \]
Of all the country-months with a positive number of reported property confiscations, 39\% have more than one reported property confiscation. The maximum is 27 reported property confiscations in a single month in the United States. Relatedly, Figure \ref{fig_confspells} documents the duration of property confiscation spells, which we define as time periods of consecutive months with a positive number of reported property confiscations within a country. 
\[ \text{Figure \ref{fig_confspells} around here.} \]
27\% of the reported property confiscations appear in consecutive months. For the United States, we observe the longest property confiscation spell, with a duration of 65 months. Overall, we observe 782 independent property confiscation spells, which corresponds to 63\% of the total number of reported property confiscations.

We define $\mbox{\textit{Confiscation}}_{imy}$ as the binary variable indicating the beginning of a property confiscation spell. We set this variable to missing in months that are part of an ongoing property confiscation spell, which started in an earlier month. Table \ref{tab_countries} in the Appendix reports the number of reported property confiscation spells per country.

%We use the GDELT data to generate additional event variables for demonstrations, strikes and boycotts, violent demonstrations, threats of any kind, and demands for a leadership change. We use these additional event variables as control variables.

%%%%%%%%%%%%%%%%%%%%%%%%%%%%%%%%%%%%%%%%%%%%%%%%%%%%%%%%%%%%%%%%%%%%%%%%%%%%%%%%%%%%
\subsection{Other data sources}

To investigate effect heterogeneity, we use GDP per capita (in current USD) from the World Development Indicators, as well as various measures of perceived corruption, government effectiveness and the rule of law. For each of these three dimensions of governance and institutional quality, we use an individual measure from the Worldwide Governance Indicators (WGIs) by \cite{kaufmann2011worldwide} as well as a commonly used alternative.\footnote{The WGIs are based on many variables provided by different organizations that all measure perceptions about some aspect of governance or institutional quality. The individual WGIs are then constructed using an unobserved component model.} We measure corruption using the WGI Control of Corruption and the Corruption Perceptions Index by Transparency International (TI). To measure government effectiveness, we use the WGI Government Effectiveness and the Quality of Government indicator by the International Country Risk Guide (ICRG), which may arguably capture a slightly broader notion of government effectiveness. We also use the Rule of Law indicators by the WGIs and Freedom House. 

%%%%%%%%%%%%%%%%%%%%%%%%%%%%%%%%%%%%%%%%%%%%%%%%%%%%%%%%%%%%%%%%%%%%%%%%%%%%%%%%%%%%
\subsection{Descriptive statistics}

Table \ref{tab_descshort} reports descriptive statistics for the two main variables for the time period 2007--2012.\footnote{Table \ref{tab_desclong} in the Appendix reports descriptive statistics for the longer sample from 2000 to 2012.}
\[ \text{Table \ref{tab_descshort} around here.} \]
Panel A includes all 160 countries. In an average country and month, the probability that a property confiscation spell starts is 7\%, and the probability that at least one offshore entity is incorporated is 27\%. Panel B excludes all countries that were classified as tax havens by \cite{hin10} or \cite{joh14}. We see that beginnings of property confiscation spells are more common, but incorporations of offshore entities are considerably less common in the sample of non-tax haven countries.

%%%%%%%%%%%%%%%%%%%%%%%%%%%%%%%%%%%%%%%%%%%%%%%%%%%%%%%%%%%%%%%%%%%%%%%%%%%%%%%%%%%%
\section{Empirical specification}

Our first objective is to estimate the effect of the beginning of a property confiscation spell on the probability that an offshore entity is incorporated in the same country in the same month. We are concerned about reverse causality and omitted variables that confound the relationship between the two. To account for these concerns, we estimate the following linear probability model:
\begin{equation}
\label{eq_emp}
\mbox{\textit{Offshore}}_{imy} = \alpha_{iy} + \beta \, \mbox{\textit{Confiscation}}_{imy} + \varepsilon_{imy}.
\end{equation}
Subscripts $i$, $m$, and $y$ indicate the country, the month and the calendar year, respectively. The parameter of main interest is $\beta$. We exploit within-country and -year variation to identify this parameter. The country-year dummies $\alpha_{iy}$ control, in the most flexible way, for all confounders that are constant within a calendar year and country. This includes all country-level variables that are available on a yearly granularity, such as measures of GDP, population, institutional quality, governance, tax revenues, etc. %Vector $X_{it}$ includes confounders that are available on a monthly granularity. We assume all variables which jointly influence the within calendar year and country variation of the endogenous and outcome variables are included in the vector $X_{it}$. This includes monthly dummies for demonstrations, strikes or boycotts, violent demonstrations, threats, and demands for a leader change. 
In other specifications, we further include month dummies $\gamma_m$ to control for seasonal effects and the lagged dependent variable to account for unobserved within-year and -country variation. The error term $\varepsilon_{imy}$ absorbs unexplained variations in offshore entity incorporations. We cluster the standard errors of the estimated coefficients at the country level. 

The specification above looks to identify a contemporaneous effect. In other words, our conjecture is that news reports on confiscations and expropriations lead to the incorporation of offshore companies within the same month. The rationale for this is the ease and speed with which offshore companies can be incorporated. Many agents provide their services publicly on the Internet. The process of advising an agent to set up an offshore company takes about ten minutes. Within a couple of days, the agent prepares the documents and registers the incorporation.   

%One possible concern of our approach is that the agents who register an offshore entity may not be located in the same country as their clients. For example, a Swiss bank may register an entity on the British Virgin Islands for a German client. This difference in locations would break the connection between a reported property confiscation within a specific country (say Germany) and the incorporation of an offshore entity from an agent from this country. In general, such a scenario most likely involves agents located in offshore countries known as tax havens. Therefore, we drop in most specifications all the countries classified as tax havens by \cite{hin10}, \cite{joh14}, and \cite{kol15}.

While the fixed effects and lagged-dependent variables control for unobserved heterogeneity, reverse causality could still be a concern. One could think of a scenario where an individual or an organization incorporates an offshore entity, which then leads to a government expropriating assets at home to either remedy the tax avoidance and evasion committed by this act or in the case of a ``bad'' government to expropriate assets not moved yet.  Our specification exploits the different time frames for the incorporation of an offshore entity and the expropriation of assets. For reverse causality to cause our results, confiscation or expropriation would need to be reported on in the same month as the incorporation that reversely caused the confiscation or expropriation. 

Confiscation processes typically take a long time. First, the incorporation has to be detected and confirmed. This typically, if it happens at all, takes some time, well beyond a month, as the incorporation of an offshore entity and the person behind it, are not observed by enforcement agencies. One major attraction of registering a company offshore is secrecy. The two main target countries identified in the Panama Papers, the British Virgin Islands and Panama itself, are in the top 20 of the current Financial Secrecy Index \citep{TJN2018,Cobham2015} with strict privacy laws.\footnote{Both destinations allow the use of so-called bearer shares. The physical holder of the bearer shares controls the company, without the owner's name being recorded in a shareholder register or anywhere else.} For a government to detect the relation of an entity and a local person or firm controlling it, in most cases a leak is required. For the period under investigation, there were no systematic large-scale leaks (such as the leaking of the Panama Papers). Moreover, just finding out about the incorporation in the same month is not enough for potential reverse causality. The actual confiscation or expropriation process has to be at least so far evolved in the same month that it is featured in the news.  Typically, a lengthy legal process is necessary before confiscation or expropriation occurs.\footnote{See, e.g., \cite{State2018,Transparency2015} and \cite{Bart2010} for descriptions of the process in the United States, the European Union and Australia, respectively.} Thus, our specification practically rules out reverse causality.

%Accordingly, the empirical specification enables us to identify contemporaneous effects of property confiscations on offshore entity incorporations, but not the (arguably even more important) long-term relationship. Seeking the long-term relationship is empirically more challenging than the contemporaneous effects, because this relationship could be blurred by reverse causality.

Our second objective is to investigate effect heterogeneity. The key question is whether the beginning of reports on property confiscations mainly leads to the incorporation of offshore entities in countries with relatively honest and effective governments or in countries with weak governance. In the former case, $\beta>0$ is consistent with the notion that shady figures use offshore entities to protect illegally acquired wealth from reasonably well-intended and well-functioning government agencies that have become more serious about enforcing the rules. In the latter case, $\beta>0$ is consistent with the notion that even honest individuals use offshore entities to protect their wealth from a Leviathan government that expropriates successful firms and individuals. To address this question, we split the sample along various measures of governance (and economic development) before estimating equation (\ref{eq_emp}) for the respective subsamples.

%%%%%%%%%%%%%%%%%%%%%%%%%%%%%%%%%%%%%%%%%%%%%%%%%%%%%%%%%%%%%%%%%%%%%%%%%%%%%%%%%%%%
\section{Empirical findings}
%%%%%%%%%%%%%%%%%%%%%%%%%%%%%%%%%%%%%%%%%%%%%%%%%%%%%%%%%%%%%%%%%%%%%%%%%%%%%%%%%%%%
\subsection{Main results}

Table \ref{tab_mainshort} documents our main estimation results.
\[ \text{Table \ref{tab_mainshort} around here.} \]
Panel A uses all observations from the sample period 2007--2012. In column (1), we report the results of a binary ordinary least squares regression. We find a positive correlation between the beginning of a property confiscation spell and the probability of the incorporation of an offshore entity. However, this correlation is not statistically significant. 

We sequentially increase the complexity of the model. In column (2), we include year and country fixed effects. In column (3), we include country-year fixed effects to estimate specification \ref{eq_emp}. This enables us to account for unobserved variables that are constant within a calendar year but may vary across calendar years within countries (such as measures for macroeconomic and political developments). The estimated coefficients are positive and statistically significant in these two columns, and very similar in magnitude. They suggest that the beginning of a property confiscation spell increases the probability of the incorporation of an offshore entity by 3 percentage points. 

We then add month-fixed effects in column (4) and the lagged dependent variable as a control in column (5). The estimated coefficient drops by around half a percentage point but is still statistically significant at the 5\% level in the most demanding specification reported in column (5). 

In Panels B, C and D of Table \ref{tab_mainshort}, we report the estimation results we drop countries classified as tax havens by \cite{hin10,joh14} or either of the two studies, respectively. Although the simple binary estimates in column (1) are quite strongly affected by the omission of tax havens, the estimates in all the other columns remain very similar. This pattern suggests that our identification strategy can account for a large amount of between-country and between-year unobserved heterogeneity. More importantly, these results show that the positive effect of the beginning of news reports on property confiscations on the incorporation of offshore entities is not driven by the (mostly small) tax havens. 

Table \ref{tab_mainlong} in the Appendix reports the estimation results for the time period 2000--2012. The results are qualitatively similar, but the size of the effects is somewhat smaller.

%%%%%%%%%%%%%%%%%%%%%%%%%%%%%%%%%%%%%%%%%%%%%%%%%%%%%%%%%%%%%%%%%%%%%%%%%%%%%%%%%%%%
\subsection{Effect heterogeneity}

We now look at heterogeneous effects and start by splitting the sample of countries that have not been classified as tax havens depending on whether their GDP per capita was above or below the median in 2006 (i.e., before our sample period). The results are reported in Table \ref{tab_hetero_gdp}.
\[ \text{Table \ref{tab_hetero_gdp} around here.} \]
The coefficient estimates in columns (2)--(5) suggest that the beginning of a property confiscation spell increases the probability of the incorporation of an offshore entity by around four percentage points in relatively rich countries but by only around one percentage point in relatively poor countries. Most of the estimates for the relatively rich countries are statistically significant, while the estimates for the relatively poor countries are not statistically significant.

We now turn to the question whether the beginning of news reports on property confiscations mainly leads to the incorporation of offshore entities in countries with good governance and sound institutions or in weakly institutionalized countries. For that purpose, we split the sample of non-tax haven countries depending on whether their level of perceived corruption was above or below the median in 2006. We do so in Table \ref{tab_hetero_cc} using the  WGI Control of Corruption in panels A and B, and the TI Corruption Perceptions Index in panels C and D. 
\[ \text{Table \ref{tab_hetero_cc} around here.} \]
The coefficient estimates in columns (2)--(5) suggest that the beginning of a property confiscation spell increases the probability of the incorporation of an offshore entity by three to six percentage points in the countries with relatively low levels of corruption (panels A and C). These estimates are mostly not statistically significant when using the WGI Control of Corruption to split the sample but are statistically significant when using TI's Corruption Perceptions Index. The effects are much smaller in magnitude and not statistically significant for the relatively corrupt countries (panels B and D). 

In Table \ref{tab_hetero_ge}, we split the sample depending on how effective the governments are using the WGI Government Effectiveness in panels A and B, and the ICRG's Quality of Government Indicator in panels C and D.
\[ \text{Table \ref{tab_hetero_ge} around here.} \]
The coefficient estimates in columns (2)--(5) suggest that the beginning of a property confiscation spell leads to a statistically significant increase in the probability of the incorporation of an offshore entity by five to seven percentage points in countries with relatively effective governments (panels A and C). There are no statistically significant effects in countries with relatively ineffective governments. 

Finally, in Table \ref{tab_hetero_rl}, we split the sample depending on the rule of law using the WGI Rule of Law in panels A and B, and the by Freedom House (FH) Rule of Law measure in panels C and D.
\[ \text{Table \ref{tab_hetero_rl} around here.} \]
The sample split based on the WGI suggests a large and statistically significant effect for countries with relatively strong rule of law but no statistically significant effect for countries with relatively weak rule of law. The sample split based on the FH measure results in a considerably smaller difference across the two sub-samples.

The pattern emerging from Tables \ref{tab_hetero_cc}--\ref{tab_hetero_rl} is consistent with the notion that shady figures use offshore entities to protect illegally acquired wealth from relatively well-intended and well-functioning governments that have become more serious about enforcing the rules. However, our results provide no evidence for the notion that individuals use offshore entities to protect their wealth from Leviathan governments that start expropriating firms and individuals more aggressively. This absence of evidence suggests that either this latter use of offshore entities is rare, or that our data and our approach fail to capture it. There are reasons to believe that the latter could be the case. For example, the GDELT Project might be less successful in extracting information from the relevant news sources in the (mostly developing) countries with Leviathan governments. Moreover, in some countries, the belief that the government is Leviathan and likely to extort successful individuals who are not part of its inner circle may already be close to one. In response to news reports on expropriations or property confiscations, these individuals would neither update their beliefs nor change their course of action. In addition, individuals from these countries may rely on agents located in tax havens rather than their own country to incorporate offshore entities. 

%%%%%%%%%%%%%%%%%%%%%%%%%%%%%%%%%%%%%%%%%%%%%%%%%%%%%%%%%%%%%%%%%%%%%%%%%%%%%%%%%%%%
\section{Conclusions}

We have argued that news reports on expropriations and property confiscations may increase the beliefs of some private individuals and firms that the government will expropriate them and confiscate their assets unless they hide them in an offshore entity. Using the Panama Papers, we have shown that the beginning of media reporting on expropriations and property confiscations in a country indeed increases the probability that offshore entities are incorporated by agents from the same country in the same month. We have documented that this effect is mainly driven by countries with relatively honest and effective governments. These findings suggest that offshore entities are used to hide wealth as honest and effective governments become more serious about enforcing the rules.

The policy debate on how to deal with tax havens has been strongly influenced by the large literature showing that individuals and firms hide wealth in offshore entities to evade (and avoid) taxes. The focus on tax evasion (and avoidance) makes sense on many grounds. Nevertheless, policy makers should not neglect that offshore entities are also used to hide wealth from honest and effective governments that become more serious about law enforcement.

%\textcolor{red}{[AS: One additional conclusion could be, that there are indirect costs of reporting property confiscation in the media, i.e., more money migrates to tax havens. This does not mean that the media should stop reporting property confiscations, but we should be aware of these negative effects.]}

%%%%%%%%%%%%%%%%%%%%%%%%%%%%%%%%%%%%%%%%%%%%%%%%%%%%%%%%%%%%%%%%%%%%%%%%%%%%%%%%%%%%
\bibliographystyle{aer}
\bibliography{bibliothek}

%%%%%%%%%%%%%%%%%%%%%%%%%%%%%%%%%%%%%%%%%%%%%%%%%%%%%%%%%%%%%%%%%%%%%%%%%%%%%%%%%%%%
\newpage
\section*{Figures and Tables}

\newpage 

\newpage 

\newpage

\begin{table}[h!]
\begin{center}
\caption{Descriptive statistics, 2007-2012.} \label{tab_descshort}
 \begin{tabular}{lccccc}
\hline
\hline
Variable & Observations & Mean &  Std.\ Dev. & Min. & Max. \\
\hline
\multicolumn{6}{l}{Panel A: Full sample} \\
\hline
$\mbox{\textit{Offshore}}_{imy}$ &  11,063 & 0.2662 & 0.4420 & 0 & 1 \\
$\mbox{\textit{Confiscation}}_{imy}$ &  11,063 & 0.0707 & 0.2563 & 0 & 1 \\
\hline
\multicolumn{6}{l}{Panel B: Tax havens dropped} \\
\hline
$\mbox{\textit{Offshore}}_{imy}$ & 7,268 & 0.1878 & 0.3906 & 0 & 1 \\
$\mbox{\textit{Confiscation}}_{imy}$ & 7,268 & 0.0951 & 0.2933 & 0 & 1 \\
\hline
\hline
\end{tabular}
\end{center}
\parbox{\textwidth}{\footnotesize Note: Time period is 2007--2012. Panel B excludes countries classified as tax havens by Hines (2010) or Johannesen and Zucman (2014; see Table \ref{tab_taxhavens} in the Appendix for details). $\mbox{\textit{Confiscation}}_{imy}$ is a binary variable indicating the beginning of a spell of months with news reports on expropriations and property confiscations in country $i$; and $\mbox{\textit{Offshore}}_{imy}$ is a binary variable indicating the incorporation of at least one offshore entity by agents from country $i$ (see Section 3 for details).}
\end{table}

\newpage

\begin{table}[h!]
\begin{center}
\caption{Main results.} \label{tab_mainshort}
 \begin{tabular}{lccccc}
\hline
\hline
% & \multicolumn{6}{c}{Outcome: Monthly dummy for offshore entity incorporation} \\
  &        (1) &        (2) &        (3) &       (4)  &        (5) \\
\hline
\multicolumn{6}{c}{Panel A: Full sample} \\
\hline
$\mbox{\textit{Confiscation}}_{imy}$  & 0.047 & 0.028** & 0.030** & 0.023* & 0.024** \\
   & (0.036) & (0.013) & (0.013) & (0.012) & (0.012) \\
Observations & 11,063 & 11,063 & 11,063 & 11,063 & 11,063 \\
\hline
\multicolumn{6}{c}{Panel B: Hines' tax havens dropped} \\
\hline
$\mbox{\textit{Confiscation}}_{imy}$  & 0.109*** & 0.027** & 0.029** & 0.022* & 0.024* \\
 & (0.035) & (0.014) & (0.014) & (0.013) & (0.013) \\
Observations & 7,910 & 7,910 & 7,910 & 7,910 & 7,910 \\
\hline
\multicolumn{6}{c}{Panel C: Johannesen and Zucman's tax havens dropped} \\
\hline
$\mbox{\textit{Confiscation}}_{imy}$  & 0.112*** & 0.033** & 0.034** & 0.027** & 0.029** \\
 & (0.034) & (0.014) & (0.014) & (0.013) & (0.013) \\
Observations & 7,625 & 7,625 & 7,625 & 7,625 & 7,625 \\
\hline
\multicolumn{6}{c}{Panel D: All tax havens dropped} \\
\hline
$\mbox{\textit{Confiscation}}_{imy}$  & 0.116*** & 0.029** & 0.030** & 0.024* & 0.026** \\
 & (0.034) & (0.034) & (0.034) & (0.034) & (0.020) \\
Observations & 7,268 & 7,268 & 7,268 & 7,268 & 7,268 \\
\hline
Country-fixed effects & No & Yes & No & No & No \\
Year-fixed effects & No & Yes & No & No & No \\
Country-year-fixed effects & No & No & Yes & Yes & Yes \\
Month-fixed effects & No & No & No & Yes & Yes \\
Lagged dependent variable & No & No & No & No & Yes \\
\hline
\hline
\end{tabular}
\end{center}
\parbox{\textwidth}{\footnotesize Note: Dependent variable is $\mbox{\textit{Offshore}}_{imy}$. Time period is 2007--2012. Panels B, C and D exclude countries classified as tax havens by Hines (2010), Johannesen and Zucman (2014) or any of the two studies, respectively (see Table \ref{tab_taxhavens} in the Appendix for country lists). $\mbox{\textit{Confiscation}}_{imy}$ is a binary variable indicating the beginning of a spell of months with news reports on expropriations and property confiscations in country $i$; and $\mbox{\textit{Offshore}}_{imy}$ is a binary variable indicating the incorporation of at least one offshore entity by agents from country $i$ (see Section 3 for details). Standard errors are clustered at the country level. ***, **, * indicate significance at the 1, 5 and 10\%-level, respectively.}
\end{table}

\newpage

\begin{table}[h!]
\begin{center}
\caption{Estimation results for rich and poor countries.} \label{tab_hetero_gdp}
 \begin{tabular}{lccccc}
\hline
\hline
% & \multicolumn{6}{c}{Outcome: Monthly dummy for offshore entity incorporation} \\
  &        (1) &        (2) &        (3) &       (4)  &        (5) \\
\hline
\multicolumn{6}{c}{Panel A: GDP per capita above median} \\
\hline
$\mbox{\textit{Confiscation}}_{imy}$  & 0.167*** & 0.044* & 0.046** & 0.035 & 0.037* \\
   & (0.042) & (0.023) & (0.023) & (0.021) & (0.021) \\
Observations & 3,503 & 3,503 & 3,503 & 3,503 & 3,503 \\
\hline
\multicolumn{6}{c}{Panel B: GDP per capita below median} \\
\hline
$\mbox{\textit{Confiscation}}_{imy}$  & 0.050 & 0.013 & 0.011 & 0.007 & 0.008 \\
 & (0.052) & (0.016) & (0.016) & (0.015) & (0.015)  \\
Observations & 3,550 & 3,550 & 3,550 & 3,550 & 3,550\\
\hline
Country-fixed effects & No & Yes & No & No & No \\
Year-fixed effects & No & Yes & No & No & No \\
Country-year-fixed effects & No & No & Yes & Yes & Yes \\
Month-fixed effects & No & No & No & Yes & Yes \\
Lagged dependent variable & No & No & No & No & Yes \\
\hline
\hline
\end{tabular}
\end{center}
\parbox{\textwidth}{\footnotesize Note: Dependent variable is $\mbox{\textit{Offshore}}_{imy}$. Time period is 2007--2012. Countries classified as tax havens by Hines (2010) or Johannesen and Zucman (2014) are excluded (see Table \ref{tab_taxhavens} in the Appendix for country lists). The sample split is based on 2006 values for GDP per capita (in current US-Dollars) from the World Development Indicators. $\mbox{\textit{Confiscation}}_{imy}$ is a binary variable indicating the beginning of a spell of months with news reports on expropriations and property confiscations in country $i$; and $\mbox{\textit{Offshore}}_{imy}$ is a binary variable indicating the incorporation of at least one offshore entity by agents from country $i$ (see Section 3 for details). Standard errors are clustered at the country level. ***, **, * indicate significance at the 1, 5 and 10\%-level, respectively.}
\end{table}

\newpage 

\begin{table}[h!]
\begin{center}
\caption{Estimation results for countries with low and high levels of corruption.} \label{tab_hetero_cc}
 \begin{tabular}{lccccc}
\hline
\hline
% & \multicolumn{6}{c}{Outcome: Monthly dummy for offshore entity incorporation} \\
  &        (1) &        (2) &        (3) &       (4)  &        (5) \\
\hline
\multicolumn{6}{c}{Panel A: Less corruption than the median according to WGI} \\
\hline
$\mbox{\textit{Confiscation}}_{imy}$  & 0.149*** & 0.041* & 0.043* & 0.033 & 0.035  \\
  & (0.040) & (0.022) & (0.022) & (0.021) & (0.021) \\
Observations &  3,538 & 3,538 & 3,538 & 3,538 & 3,538  \\
\hline
\multicolumn{6}{c}{Panel B: More corruption than the median according to WGI} \\
\hline
$\mbox{\textit{Confiscation}}_{imy}$   
&0.072 & 0.018 & 0.018 & 0.014 & 0.015 \\
 &  (0.053) & (0.017) & (0.017) & (0.015) & (0.016) \\
Observations & 3,514 & 3,514 & 3,514 & 3,514 & 3,514 \\
\hline
\multicolumn{6}{c}{Panel C: Less corruption than the median according to TI} \\
\hline
$\mbox{\textit{Confiscation}}_{imy}$ & 0.157*** & 0.054** & 0.056*** & 0.040** & 0.042** \\
 & (0.043) & (0.021) & (0.021) & (0.020) & (0.019) \\
Observations & 3,598 & 3,598 & 3,598 & 3,598 & 3,598 \\
\hline
\multicolumn{6}{c}{Panel D: More corruption than the median according to TI} \\
\hline
$\mbox{\textit{Confiscation}}_{imy}$ & 0.023 & -0.001 & -0.004 & -0.003 & -0.003 \\
 & (0.042) & (0.015) & (0.015) & (0.013) & (0.013) \\
Observations & 3,310 & 3,310 & 3,310 & 3,310 & 3,310 \\
\hline
Country-fixed effects & No & Yes & No & No & No \\
Year-fixed effects & No & Yes & No & No & No \\
Country-year-fixed effects & No & No & Yes & Yes & Yes \\
Month-fixed effects & No & No & No & Yes & Yes \\
Lagged dependent variable & No & No & No & No & Yes \\
\hline
\hline
\end{tabular}
\end{center}
\parbox{\textwidth}{\footnotesize Note: Dependent variable is $\mbox{\textit{Offshore}}_{imy}$. Time period is 2007--2012. Countries classified as tax havens by Hines (2010) or Johannesen and Zucman (2014) are excluded (see Table \ref{tab_taxhavens} in the Appendix for country lists). The sample split is based on 2006 values for the Worldwide Governance Indicator (WGI) Control of Corruption in Panels A and B, and the Corruption Perceptions Index by Transparency International (TI) in Panels C and D. $\mbox{\textit{Confiscation}}_{imy}$ is a binary variable indicating the beginning of a spell of months with news reports on expropriations and property confiscations in country $i$; and $\mbox{\textit{Offshore}}_{imy}$ is a binary variable indicating the incorporation of at least one offshore entity by agents from country $i$ (see Section 3 for details). Standard errors are clustered at the country level. ***, **, * indicate significance at the 1, 5 and 10\%-level, respectively.}
\end{table}

\newpage

\begin{table}[h!]
\begin{center}
\caption{Estimation results for countries with low and high government effectiveness.} \label{tab_hetero_ge}
 \begin{tabular}{lccccc}
\hline
\hline
% & \multicolumn{6}{c}{Outcome: Monthly dummy for offshore entity incorporation} \\
  &        (1) &        (2) &        (3) &       (4)  &        (5) \\
\hline
\multicolumn{6}{c}{Panel A: More effective government than the median according to WGI} \\
\hline
$\mbox{\textit{Confiscation}}_{imy}$  
& 0.177*** & 0.072*** & 0.069*** & 0.052** & 0.053** \\
 & (0.047) & (0.022) & (0.022) & (0.021) & (0.021) \\
Observations & 3,443 & 3,443 & 3,443 & 3,443 & 3,443  \\
\hline
\multicolumn{6}{c}{Panel B: Less effective government than the median according to WGI} \\
\hline
$\mbox{\textit{Confiscation}}_{imy}$ & 0.029 & -0.012 & -0.011 & -0.010 & -0.010  \\
 & (0.040) & (0.013) & (0.013) & (0.012) & (0.012)  \\
Observations & 33,609 & 3,609 & 3,609 & 3,609 & 3,609 \\
\hline
\multicolumn{6}{c}{Panel C: More effective government than the median according to ICRG} \\
\hline
$\mbox{\textit{Confiscation}}_{imy}$ & 0.148*** & 0.068*** & 0.065*** & 0.047** & 0.049**  \\
 & (0.045) & (0.020) & (0.020) & (0.020) & (0.020) \\
Observations & 3,301 & 3,301 & 3,301 & 3,301 & 3,301 \\
\hline
\multicolumn{6}{c}{Panel D: Less effective government than the median according to ICRG} \\
\hline
$\mbox{\textit{Confiscation}}_{imy}$ & 0.036 & -0.013 & -0.010 & -0.007 & -0.007 \\
 & (0.048) & (0.017) & (0.017) & (0.015) & (0.015) \\
Observations & 3,248 & 3,248 & 3,248 & 3,248 & 3,248  \\
\hline
Country-fixed effects & No & Yes & No & No & No \\
Year-fixed effects & No & Yes & No & No & No \\
Country-year-fixed effects & No & No & Yes & Yes & Yes \\
Month-fixed effects & No & No & No & Yes & Yes \\
Lagged dependent variable & No & No & No & No & Yes \\
\hline
\hline
\end{tabular}
\end{center}
\parbox{\textwidth}{\footnotesize Note: Dependent variable is $\mbox{\textit{Offshore}}_{imy}$. Time period is 2007--2012. Countries classified as tax havens by Hines (2010) or Johannesen and Zucman (2014) are excluded (see Table \ref{tab_taxhavens} in the Appendix for country lists). The sample split is based on 2006 values for the Worldwide Governance Indicator (WGI) Government Effectiveness in Panels A and B, and the Quality of Government indicator by the International Country Risk Guide (ICRG) in Panels C and D. $\mbox{\textit{Confiscation}}_{imy}$ is a binary variable indicating the beginning of a spell of months with news reports on expropriations and property confiscations in country $i$; and $\mbox{\textit{Offshore}}_{imy}$ is a binary variable indicating the incorporation of at least one offshore entity by agents from country $i$ (see Section 3 for details). Standard errors are clustered at the country level. ***, **, * indicate significance at the 1, 5 and 10\%-level, respectively.}
\end{table}

\newpage

\begin{table}[h!]
\begin{center}
\caption{Estimation results for countries with strong and weak rule of law.} \label{tab_hetero_rl}
 \begin{tabular}{lccccc}
\hline
\hline
% & \multicolumn{6}{c}{Outcome: Monthly dummy for offshore entity incorporation} \\
  &        (1) &        (2) &        (3) &       (4)  &        (5) \\
\hline
\multicolumn{6}{c}{Panel A: Stronger rule of law than the median according to WGI} \\
\hline
$\mbox{\textit{Confiscation}}_{imy}$  & 0.155*** & 0.050** & 0.050** & 0.036* & 0.038*  \\
  & (0.043) & (0.022) & (0.022) & (0.021) & (0.021) \\
Observations & 3,492 & 3,492 & 3,492 & 3,492 & 3,492  \\
\hline
\multicolumn{6}{c}{Panel B: Weaker rule of law than the median according to WGI} \\
\hline
$\mbox{\textit{Confiscation}}_{imy}$ & 0.063 & 0.010 & 0.010 & 0.011 & 0.012 \\
 & (0.052) & (0.016) & (0.016) & (0.014) & (0.014) \\
Observations & 3,560 & 3,560 & 3,560 & 3,560 & 3,560 \\
\hline
\multicolumn{6}{c}{Panel C: Stronger rule of law than the median according to FH} \\
\hline
$\mbox{\textit{Confiscation}}_{imy}$ & 0.112*** & 0.038* & 0.037* & 0.025 & 0.026 \\
  & (0.038) & (0.021) & (0.021) & (0.020) & (0.020) \\
Observations &  3,906 & 3,906 & 3,906 & 3,906 & 3,906  \\
\hline
\multicolumn{6}{c}{Panel D: Weaker rule of law than the median according to FH} \\
\hline
$\mbox{\textit{Confiscation}}_{imy}$ & 0.108* & 0.021 & 0.022 & 0.022 & 0.023 \\
 & (0.059) & (0.017) & (0.018) & (0.016) & (0.016)\\
Observations &3,146 & 3,146 & 3,146 & 3,146 & 3,146 \\
\hline
Country-fixed effects & No & Yes & No & No & No \\
Year-fixed effects & No & Yes & No & No & No \\
Country-year-fixed effects & No & No & Yes & Yes & Yes \\
Month-fixed effects & No & No & No & Yes & Yes \\
Lagged dependent variable & No & No & No & No & Yes \\
\hline
\hline
\end{tabular}
\end{center}
\parbox{\textwidth}{\footnotesize Note: Dependent variable is $\mbox{\textit{Offshore}}_{imy}$. Time period is 2007--2012. Countries classified as tax havens by Hines (2010) or Johannesen and Zucman (2014) are excluded (see Table \ref{tab_taxhavens} in the Appendix for country lists). The sample split is based on 2006 values for the Worldwide Governance Indicator (WGI) Rule of Law in Panels A and B, and the Rule of Law indicator by Freedom House (FH) in Panels C and D. $\mbox{\textit{Offshore}}_{imy}$ and $\mbox{\textit{Confiscation}}_{imy}$ is a binary variable indicating the beginning of a spell of months with news reports on expropriations and property confiscations in country $i$; and $\mbox{\textit{Offshore}}_{imy}$ is a binary variable indicating the incorporation of at least one offshore entity by agents from country $i$ (see Section 3 for details). Standard errors are clustered at the country level. ***, **, * indicate significance at the 1, 5 and 10\%-level, respectively.}
\end{table}

\newpage
	 
%%%%%%%%%%%%%%%%%%%%%%%%%%%%%%%%%%%%%%%%%%%%%%%%%%%%%%%%%%%%%%%%%%%%%%%%%%%%%%%%%%%%
\renewcommand\appendix{\par
   \setcounter{section}{0}%
   \setcounter{subsection}{0}%
   \setcounter{table}{0}%
	\setcounter{figure}{0}%
   \renewcommand\thesection{\Alph{section}}%
   \renewcommand\thetable{A.\arabic{table}}}
\renewcommand\thefigure{A.\arabic{figure}}
\clearpage
\begin{appendix}
\noindent \textbf{\Large Appendix: Additional tables}

\begin{table}[H]
\begin{center}
\caption{Jurisdictions of offshore entities in the Panama Papers.}   \label{tab_entjur}
 \begin{tabular}{lccc}
    \hline \hline
 	 \multirow{2}{*}{Jurisdiction} & Absolute & Relative & Cumulated \\
	 & frequency & frequency & frequency  \\ \hline
 Virgin Islands, British & 112,915 & 53.12\% & 53.12\% \\
	Panama & 48,214 & 22.68\% & 75.80\%   \\
	Bahamas & 15,870 & 7.466\% & 83.27\%   \\
	Seychelles & 15,133 & 7.119\% & 90.39\%   \\
	Niue & 9,567 & 4.501\% & 94.89\%   \\
	Samoa & 5,292 & 2.490\% & 97.38\%   \\
 	Anguilla & 3,232 & 1.520\% & 98.90\%   \\
	Nevada & 1,255 & 0.590\% & 99.49\%   \\
	Hong Kong & 452 & 0.213\% & 99.70\%   \\
	United Kingdom & 145 & 0.068\% & 99.77\%  \\
	Belize & 130 & 0.061\% & 99.83\%  \\
	Costa Rica & 78 & 0.037\% & 99.87\%  \\
	Cyprus & 76 & 0.036\% & 99.90\%   \\
	Uruguay & 52 & 0.024\% & 99.93\%   \\
	New Zealand & 47 & 0.022\% & 99.95\%   \\
	Jersey & 39 & 0.018\% & 99.97\%   \\
	Wyoming & 37 & 0.017\% & 99.98\%   \\
	Malta & 28 & 0.013\% & 100.00\%   \\
	Isle of Man & 7 & 0.003\% & 100.00\%   \\
	Ras al Khaimah & 2 & 0.001\% & 100.00\%   \\
	Singapore & 1 & 0.0005\% & 100.00\%  \\
    \hline
    Total & 212,572 & 100\% & \\
    \hline
    \hline
    \end{tabular}
\parbox{10.7cm}{\footnotesize Note: Own calculations based on data from the Panama Papers. }
\end{center}
\end{table}

\newpage

\begin{table}[H]
\begin{small}
\begin{center}
\caption{Countries/jurisdictions of agents' registering offshore entities.} \label{tab_countries}
\begin{tabular}{lcclcc} \hline \hline
   Agents' & Months with & Property   &    Agents' & Months with  & Property   \\
 country/ & offshore entity  & confiscation &  country/ &  offshore entity  & confiscation \\
jurisdiction  &incorporations & spells & jurisdiction & incorporations &  spells \\
\hline
    Albania & 0     & 1     & Ecuador & 60    & 4 \\
    Am. Samoa & 1     & 0     & Egypt & 3     & 10 \\
    Andorra & 35    & 0     & El Salvador & 42    & 1 \\
    Angola & 7     & 0     & Estonia & 51    & 2 \\
    Anguilla & 14    & 0     & Finland & 9     & 2 \\
    Antigua \& Barb. & 8     & 2     & France & 20    & 6 \\
    Argentina & 36    & 3     & Georgia & 0     & 0 \\
    Aruba & 10    & 0     & Germany & 18    & 16 \\
    Australia & 6     & 13    & Ghana & 1     & 8 \\
    Austria & 2     & 5     & Gibraltar & 60    & 0 \\
    Azerbaijan & 2     & 3     & Greece & 25    & 9 \\
    Bahamas & 61    & 1     & Guam  & 0     & 0 \\
    Bahrain & 0     & 4     & Guatemala & 57    & 2 \\
    Bangladesh & 2     & 13    & Guernsey & 60    & 0 \\
    Barbados & 8     & 1     & Haiti & 1     & 3 \\
    Belarus & 4     & 5     & Honduras & 5     & 5 \\
    Belgium & 6     & 0     & Hong Kong & 64    & 0 \\
    Belize & 51    & 0     & Hungary & 8     & 7 \\
    Bermuda & 1     & 2     & Iceland & 2     & 3 \\
    Bolivia & 32    & 3     & India & 6     & 17 \\
    Botswana & 0     & 1     & Indonesia & 3     & 15 \\
    Brazil & 57    & 5     & Iran  & 0     & 19 \\
    Brunei Darus. & 1     & 0     & Ireland & 16    & 11 \\
    Bulgaria & 7     & 6     & Isle of Man & 61    & 0 \\
    Cameroon & 2     & 0     & Israel & 37    & 17 \\
    Canada & 27    & 16    & Italy & 28    & 16 \\
    Cayman Islands & 17    & 1     & Jamaica & 0     & 5 \\
    Centr. Afr. Rep. & 1     & 1     & Japan & 5     & 10 \\
    Chad  & 3     & 0     & Jersey & 63    & 0 \\
    Chile & 25    & 2     & Jordan & 60    & 7 \\
    China & 46    & 18    & Kazakhstan & 0     & 2 \\
    Colombia & 56    & 13    & Kenya & 4     & 10 \\
    Cook Islands & 0     & 0     & Korea & 2     & 11 \\
    Costa Rica & 55    & 2     & Kuwait & 7     & 3 \\
    Croatia & 2     & 1     & Latvia & 44    & 1 \\
    Cuba  & 1     & 7     & Lebanon & 51    & 9 \\
    Curaçao & 6     & 0     & Lesotho & 0     & 0 \\
    Cyprus & 59    & 3     & Liberia & 0     & 7 \\
    Czech Republic & 10    & 8     & Libya & 0     & 9 \\
    Côte d'Ivoire & 4     & 1     & Liechtenst. & 60    & 0 \\
    Denmark & 2     & 0     & Lithuania & 4     & 4 \\
    Djibouti & 1     & 0     & Luxemb. & 62    & 1 \\
    Dominica & 24    & 0     & Macao & 6     & 0 \\
    Dominican Rep. & 50    & 6     & Macedonia & 0     & 2 \\
\hline
\hline
\end{tabular}
\parbox{16cm}{\centering < table continues on next page >  }
\end{center}
\end{small}
\end{table}

\newpage

\begin{table}[H]
\begin{small}
\begin{center}
\parbox{16cm}{\centering \normalsize Table \ref{tab_countries}: < continued > }
\begin{tabular}{lcclcc} \hline \hline
   Agents' & Months with  & Property   &    Agents' & Months with  & Property   \\
 country/ & offshore entity  &  confiscation &  country/ &  offshore entity  & confiscation \\
jurisdiction  &incorporations & spells & jurisdiction & incorporations &  spells \\
\hline
    Malawi & 0     & 1     & Saudi Arabia & 13    & 12 \\
    Malaysia & 3     & 12    & Senegal & 0     & 1 \\
    Mali  & 0     & 1     & Seychelles & 60    & 1 \\
    Malta & 54    & 2     & Singapore & 61    & 5 \\
    Marshall Is. & 0     & 0     & Sint Maarten  & 0     & 0 \\
    Mauritius & 59    & 0     & Slovakia & 0     & 3 \\
    Mexico & 14    & 15    & Slovenia & 4     & 0 \\
    Moldova & 0     & 0     & South Africa & 12    & 5 \\
    Monaco & 60    & 1     & Spain & 46    & 12 \\
    Montenegro & 5     & 0     & Sri Lanka & 0     & 6 \\
    Morocco & 13    & 3     & Sudan & 0     & 13 \\
    Mozambique & 0     & 1     & Sweden & 2     & 3 \\
    Namibia & 1     & 2     & Switzerland & 59    & 8 \\
    Nauru & 1     & 0     & Syria & 0     & 8 \\
    Netherlands & 18    & 6     & Taiwan & 50    & 7 \\
    New Zealand & 7     & 9     & Tanzania & 0     & 5 \\
    Nicaragua & 1     & 3     & Thailand & 46    & 12 \\
    Nigeria & 2     & 16    & Trin. \& Tob. & 1     & 2 \\
    Niue  & 0     & 0     & Tunisia & 0     & 10 \\
    Norway & 0     & 5     & Turkey & 11    & 14 \\
    Oman  & 2     & 1     & Turks \& C. Is. & 2     & 0 \\
    Pakistan & 0     & 13    & Uganda & 0     & 11 \\
    Panama & 62    & 2     & Ukraine & 43    & 9 \\
    Paraguay & 19    & 2     & Un. Arab Em. & 62    & 11 \\
    Peru  & 12    & 3     & Ukraine & 43    & 18 \\
    Philippines & 2     & 12    & US    & 5     & 3 \\
    Poland & 13    & 7     & Uruguay & 61    & 0 \\
    Portugal & 16    & 2     & Uzbekistan & 2     & 5 \\
    Puerto Rico & 0     & 0     & Vanuatu & 0     & 0 \\
    Qatar & 4     & 1     & Venezuela & 53    & 13 \\
    Romania & 1     & 10    & Vietnam & 2     & 5 \\
    Russian Fed. & 40    & 16    & Virgin Is., Br. & 58    & 0 \\
    St. Kitts \& N. & 31    & 0     & Virgin Is., US & 1     & 0 \\
    St. Lucia & 0     & 0     & Yemen & 0     & 10 \\
    St. Vinc. \& G. & 1     & 0     & Zambia & 0     & 11 \\
    Samoa & 60    & 0     & Zimbabwe & 0     & 12 \\
\hline
\hline
\end{tabular}
\parbox{15cm}{\footnotesize Note: Own calculations based on Panama Papers and GDELT data for the years 2007-2012.}
\end{center}
\end{small}
\end{table}

\newpage

\begin{table}[H]
\begin{center}
\caption{List of countries/jurisdictions classified as tax havens.} \label{tab_taxhavens}
\begin{footnotesize}
\begin{tabular}{lcc}
\hline \hline
\multirow{2}{*}{Country/jurisdiction} & Hines & Johannesen and \\ 
 & (2010) & Zucman (2010) \\
\hline
Andorra & x & x \\ 
Anguilla & x & x \\ 
Antigua and Barbuda & x & x \\ 
Aruba & x & x \\ 
Austria &  & x \\ 
Bahamas & x & x \\ 
Bahrain & x & x \\ 
Barbados & x & x \\ 
Belgium &  & x \\ 
Belize & x & x \\ 
Bermuda & x & x \\ 
Cayman Islands & x & x \\ 
Chile &  & x \\ 
Cook Islands & x & x \\ 
Costa Rica & x & x \\ 
Curacao &  & x \\ 
Cyprus & x & x \\ 
Djibouti & x & \\
Dominica & x & x \\ 
Gibraltar & x & x \\  
Guernsey & x & x \\ 
Hong Kong & x & x \\ 
Ireland & x & \\        
Isle of Man & x & x \\ 
Jersey & x & x \\ 
Jordan & x & \\
Lebanon & x & \\
Liberia & x & x \\ 
Liechtenstein & x & x \\ 
Luxembourg & x & x \\ 
Macao & x & x \\ 
Malaysia &  & x \\ 
Malta & x & x \\ 
Marshall Islands & x & x \\ 
Mauritius & x & \\
Monaco & x & x \\ 
Nauru & x & x \\ 
Niue & x & x \\ 
Panama & x & x \\ 
Saint Kitts and Nevis & x & x \\ 
Saint Lucia & x & x \\ 
Saint Vincent and the G. & x & x \\ 
Samoa & x & x \\ 
Seychelles & x & x \\ 
Singapore & x & x \\ 
Sint Maarten &  & x \\ 
Switzerland & x & x \\ 
Trinidad and Tobago &  & x \\ 
Turks and Caicos Islands & x & x \\ 
Uruguay &  & x \\ 
Vanuatu & x & x \\ 
Virgin Islands, British & x & x \\ 
Virgin Islands, U.S. &  & x \\ 
\hline
\end{tabular}
\parbox{8.2cm}{Note: This list is based on Table A.1 in \cite{kol15}.}
\end{footnotesize}
\end{center}
\end{table}

\newpage

\begin{table}[h!]
\begin{center}
\caption{Descriptive statistics for time period 2000--2012.} \label{tab_desclong}
 \begin{tabular}{lccccc}
\hline
\hline
Variable & Observations & Mean &  Std.\ Dev. & Min. & Max. \\
\hline
\multicolumn{6}{l}{Panel A: Full sample} \\
\hline
$\mbox{\textit{Offshore}}_{imy}$ &  24,406 & 0.2655 & 0.4416 & 0 & 1 \\
$\mbox{\textit{Confiscation}}_{imy}$ &  24,406 & 0.0497 & 0.2172 & 0 & 1 \\
\hline
\multicolumn{6}{l}{Panel B: Tax havens dropped} \\
\hline
$\mbox{\textit{Offshore}}_{imy}$ & 16,161 & 0.1880 & 0.3907 & 0 & 1 \\
$\mbox{\textit{Confiscation}}_{imy}$ & 16,161 & 0.0674 & 0.2507 & 0 & 1 \\
\hline
\hline
\end{tabular}
\end{center}
\parbox{\textwidth}{\footnotesize Note: Time period is 2000--2012. Panel B excludes countries classified as tax havens by Hines (2010) or Johannesen and Zucman (2014; see Table \ref{tab_taxhavens} in the Appendix for details). $\mbox{\textit{Confiscation}}_{imy}$ is a binary variable indicating the beginning of a spell of months with news reports on expropriations and property confiscations in country $i$; and $\mbox{\textit{Offshore}}_{imy}$ is a binary variable indicating the incorporation of at least one offshore entity by agents from country $i$ (see Section 3 for details).}
\end{table}

\newpage

\begin{table}[h!]
\begin{center}
\caption{Estimation results for the time period 2000--2012.} \label{tab_mainlong}
 \begin{tabular}{lccccc}
\hline
\hline
% & \multicolumn{6}{c}{Outcome: Monthly dummy for offshore entity incorporation} \\
  &        (1) &        (2) &        (3) &       (4)  &        (5) \\
\hline
\multicolumn{6}{c}{Panel A: Full sample} \\
\hline
$\mbox{\textit{Confiscation}}_{imy}$  & 0.048 & 0.019** & 0.017* & 0.014 & 0.015* \\
   & (0.036) & (0.009) & (0.010) & (0.009) & (0.009) \\
Observations & 24,406 & 24,406 & 24,406 & 24,406 & 24,406 \\
\hline
\multicolumn{6}{c}{Panel B: Hines' tax havens dropped} \\
\hline
$\mbox{\textit{Confiscation}}_{imy}$  & 0.109*** & 0.021** & 0.018* & 0.015 & 0.016* \\
 & (0.034) & (0.009) & (0.010) & (0.009) & (0.009) \\
Observations & 17,559 & 17,559 & 17,559 & 17,559 & 17,559 \\
\hline
\multicolumn{6}{c}{Panel C: Johannesen and Zucman's tax havens dropped} \\
\hline
$\mbox{\textit{Confiscation}}_{imy}$  & 0.107*** & 0.023** & 0.022** & 0.018** & 0.019** \\
 & (0.033) & (0.009) & (0.010) & (0.009) & (0.009) \\
Observations & 16,938 & 16,938 & 16,938 & 16,938 & 16,938 \\
\hline
\multicolumn{6}{c}{Panel D: All tax havens dropped} \\
\hline
$\mbox{\textit{Confiscation}}_{imy}$  & 0.114*** & 0.022** & 0.019* & 0.016* & 0.017* \\
 & (0.034) & (0.009) & (0.010) & (0.009) & (0.009) \\
Observations & 16,161 & 16,161 & 16,161 & 16,161 & 16,161 \\
\hline
Country-fixed effects & No & Yes & No & No & No \\
Year-fixed effects & No & Yes & No & No & No \\
Country-year-fixed effects & No & No & Yes & Yes & Yes \\
Month-fixed effects & No & No & No & Yes & Yes \\
Lagged dependent variable & No & No & No & No & Yes \\
\hline
\hline
\end{tabular}
\end{center}
\parbox{\textwidth}{\footnotesize Note: Dependent variable is $\mbox{\textit{Offshore}}_{imy}$. Time period is 2000--2012. Panels B, C and D exclude countries classified as tax havens by Hines (2010), Johannesen and Zucman (2014) or any of the two studies, respectively (see Table \ref{tab_taxhavens} in the Appendix for country lists). $\mbox{\textit{Confiscation}}_{imy}$ is a binary variable indicating the beginning of a spell of months with news reports on expropriations and property confiscations in country $i$; and $\mbox{\textit{Offshore}}_{imy}$ is a binary variable indicating the incorporation of at least one offshore entity by agents from country $i$ (see Section 3 for details). Standard errors are clustered at the country level. ***, **, * indicate significance at the 1, 5 and 10\%-level, respectively.}
\end{table}

\end{appendix}
\end{document}